\documentclass[9pt,twocolumn,twoside]{optica}
\usepackage{textgreek}
\usepackage{xcolor}
\usepackage{braket}
\setboolean{shortarticle}{false}
\setboolean{minireview}{false}
\usepackage{flushend}
\usepackage[normalem]{ulem}

\title{All-optical phase-sensitive detection for ultra-fast quantum computation}

\author[1]{Naoto Takanashi}
\author[2]{Asuka Inoue}
\author[2]{Takahiro Kashiwazaki}
\author[2]{Takushi Kazama}
\author[2]{Koji Enbutsu}
\author[2]{Ryoichi Kasahara}
\author[2]{Takeshi Umeki}
\author[1,*]{Akira Furusawa}

\affil[1]{Department of Applied Physics, School of Engineering, The University of Tokyo, 7-3-1 Hongo, Bunkyo-ku, Tokyo, 113-8656, Japan}
\affil[2]{NTT Device Technology Labs, NTT Corporation, 3-1, Morinosato Wakamiya, Atsugi, Kanagawa, 243-0198, Japan}
\affil[*]{Corresponding author: akiraf@ap.t.u-tokyo.ac.jp}

\begin{abstract}
Phase-sensitive detection is the essential projective measurement for measurement-based continuous-variable quantum information processing. The bandwidth of conventional electrical phase-sensitive detectors is up to several gigahertz, which would limit the speed of quantum computation. It is theoretically proposed to realize terahertz-order detection bandwidth by using all-optical phase-sensitive detection with an optical parametric amplifier (OPA). However, there have been experimental obstacles to achieve large parametric gain for continuous waves, which is required for use in quantum computation. Here, we adopt a fiber-coupled $\chi^{(2)}$ OPA made of a periodically poled LiNbO${}_{3}$ waveguide with high durability for intense continuous-wave pump light. Thanks to that, we manage to detect quadrature amplitudes of broadband continuous-wave squeezed light. 3 dB of squeezing is measured up to 3 THz of sideband frequency with an optical spectrum analyzer. Furthermore, we demonstrate the phase-locking and dispersion compensation of the broadband continuous-wave squeezed light, so that the phase of the squeezed light is maintained over 1 THz. The ultra-broadband continuous-wave detection method and dispersion compensation would help to realize all-optical quantum computation with over-THz clock frequency.
\end{abstract}

\setboolean{displaycopyright}{false}

\begin{document}

\maketitle
\section{Introduction}
\begin{figure*}[!t]
\centering
\includegraphics[width=18.3cm]{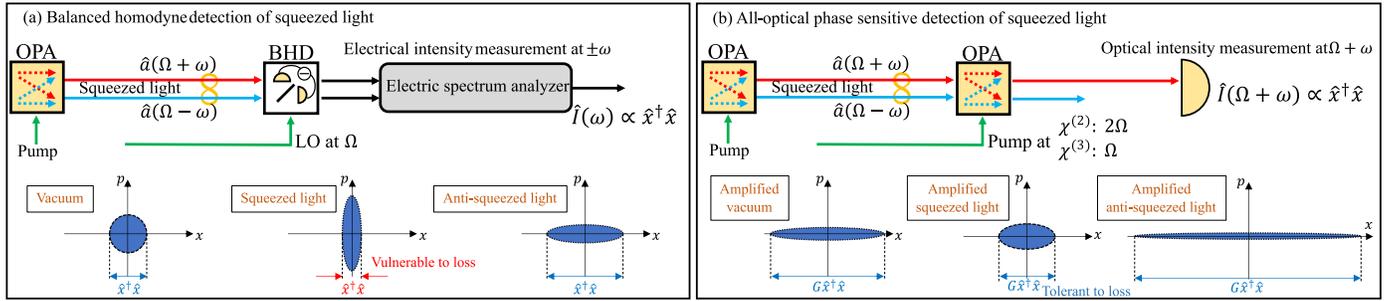}
\caption{Schematic of phase-sensitive detection with (a) a conventional balanced homodyne detector (b) an optical parametric amplifier (OPA). In the both cases, the measured light is squeezed light generated in an OPA. Squeezed light has quantum entanglements between the sidebands, and phase-sensitive detection provides a way to measure the strength of this correlation, namely squeezing level. Squeezed quadrature amplitude is especially vulnerable to optical loss. Parametric amplification along the measurement axis anti-squeezes the measured quadrature, and improves the tolerance to optical loss. $\Omega$ is the center frequency of the OPAs, and $\omega$ is sideband frequency. BHD, balanced homodyne detector.}
\label{concept}
\end{figure*}

All-optical quantum computation is an ultimate goal of the research of quantum information processing which pursues speed of computation. Generation, detection, and feed-forward control of a multipartite entangled state such as a two-dimensional cluster state \cite{twoDcluster2,twoDcluster} are the essential processes of measurement-based quantum computation \cite{OneWay,MillionModes:Yoshikawa}.
The speed of quantum computation is determined by the bandwidths of the generated entangled state, detection and feed-forward control. As for the entangled state, its bandwidth is inherited from the squeezed light used for its generation. Squeezed light has non-classical correlation between its sidebands, and the bandwidth of squeezed light recently reaches 2 THz \cite{PPLNOPA6dB}. However, the bandwidth of detection and feed-forward control would be limited up to several GHz as long as electrical circuits are used. Here, the all-optical scheme aims to break the limitation by replacing the electrical circuits with nonlinear optical elements.

In 1999, an all-optical implementation of quantum teleportation, which is the simplest case of measurement-based quantum computation, is theoretically proposed \cite{alloptical}. Here, optical parametric amplifiers (OPAs) made of nonlinear media play a role in converting quantum field of light into loss-tolerant ``classical'' field, which can be directly used for feed-forward control. Meanwhile, in conventional non-all-optical scheme, quantum field is converted into electrical signal with balanced homodyne detection, and then the signal is used for modulation for feed-forward control of another light \cite{FurusawaTeleportation}.

Balanced homodyne detection is phase-sensitive detection in which signal light and a local oscillator are mixed by an optical beamsplitter and converted into a low-frequency electrical signal by balanced photodiodes. For the measurement of squeezing level, the electrical signal is conventionally detected by an electrical spectrum analyzer as in Fig. \ref{concept}(a), where one can get information from both sidebands simultaneously and obtain their non-classical correlation, namely squeezing level \cite{Yuen1,Yuen2,EPRinSqueezing}. Here, the quadrature amplitude is vulnerable to optical loss, and it is required for photodiodes used in homodyne detection to have nearly 100\% quantum efficiency. Photodiodes with high quantum efficiency and low electrostatic capacitance have been developed and homodyne detection with the gigahertz-order bandwidth has been achieved \cite{GHzSq:Schnabel,FastHomodyne1,FastHomodyne2}. However, broadening bandwidth further might accompany decrease in quantum efficiency \cite{serikawatheory,serikawadetector}.

Since the bandwidth of an OPA can reach several terahertz, \cite{BroadbandOPA,FiberBroadbandOPA,WaveguideBroadbandOPA,OPAFurusawa}, the utilization of OPAs for detection is promising to speed up quantum information processing. Theoretical proposals have been made to utilize parametric amplification for measurement of light so far \cite{YamamotoQND,NoiseFactorOfAmplifiers,NoiseFactorOfAmplifiers2,SU(2)andSU(11)}. Especially, the measurement of light with non-classical correlation is important for quantum computation. In 2020, a measurement method for Einstein-Podolsky-Rosen-type entanglement using an OPA and a low-quantum-efficiency receiver was proposed \cite{EntangledetectionbyOPA}. Parametrically amplified light is tolerant to optical loss \cite{LossToleranceOfSU11}, so that it can be converted into electrical signals with broadband low-quantum-efficiency detector \cite{EntangledetectionbyOPA} or directly used for all-optical feed-forward control \cite{alloptical,allopticalexperiment}. For the implementation of this technique, an OPA should have large gain for continuous waves, but it is challenging because it requires high durability for an intense pump beam \cite{Kashiwazaki}.

In 2018, the measurement of squeezing level with parametric amplification was achieved using a $\chi^{(3)}$ photonic crystal fiber pumped by a pulsed laser \cite{LiftingBandwidth}. Here, the pulsed pump is used to attain momentary large parametric gain at its peak. However, it could be a problem that the pulsed pump limits the operation time of quantum computing. To measure time-domain-multiplexed quantum states, a continuous operation is an indispensable feature because it can always accept input light, while a pulsed operation accepts input light only at the moment of pumping. Moreover, the bandwidth of the measured squeezed light was limited by chromatic dispersion in the experiment, which might be an obstacle to realize ultra-fast quantum information processing. In addition, $\chi^{(3)}$ nonlinearity requires intense pump light at the center wavelength of the signal, which could cause unwanted nonlinear effects, such as self-phase modulation or cross-phase modulation, and difficulties in separation of the pump light. Here, in $\chi^{(2)}$ parametric amplification, the wavelength of the pump light is a half of that of the signal, and they can be easily separated by dichroic mirrors \cite{4dBFiberedOPA}. In 2019, a highly durable $\chi^{(2)}$ OPA made of a periodically poled LiNbO${}_{3}$ (PPLN) waveguide was developed and the gain up to 30 dB for continuous-wave input light was achieved with over-one-watt continuous-wave pump light \cite{Kashiwazaki}.

In this letter, we demonstrate ultra-broadband detection of a continuous-wave squeezed light with a single-mode PPLN waveguide \cite{PPLNOPA6dB}. The waveguide OPA is connected to another waveguide OPA for squeezed-light generation with single-mode polarization-maintaining (PM) fibers. The fiber-coupled input would be compatible with integrated quantum circuits in the future. 3 dB of squeezing is observed over 3 THz sideband frequency with an optical spectrum analyzer. Here, phase locking of the squeezed light is performed with a variable bandpass fiter, which allows to lock the phase to an arbitrary point. 
Furthermore, we demonstrate dispersion compensation of the broadband squeezed light, and the phase of the squeezed light is maintained over 1 THz. Our work would contribute to the realization of all-optical quantum computing with over-THz clock frequency.
%%%%%%%%%%%%%%%%%%%%%%%%%%%%%%%%%%%%%%%%%%%%%%%%%%%%%%%%%%%%%%%%%%%%%%%%%%%%%%%%
\section{All-optical phase-sensitive detection of a squeezed vacuum}
\begin{figure*}[!t]
\centering\includegraphics[width=18.3cm]{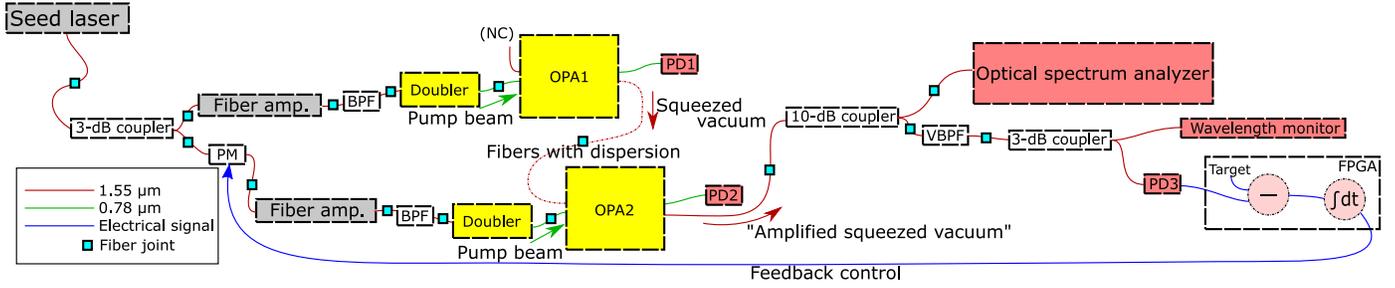}
\caption{A schematic of the experimental setup. The two OPAs are pumped by using separate frequency doublers and fiber amplifiers. The output of the second OPA is split into two beams. One is used for phase locking, and the other is measured by an optical spectrum analyzer. BPF, band-pass filter; PM, phase modulator; PD, photo detector; VBPF, variable band-pass filter; FPGA, field-programmable gate array.}
\label{setup}
\end{figure*}
Quantum entanglement between each sideband is the essential feature of squeezed light. Phase-sensitive detection measures the strength of the quantum correlation, namely squeezing level \cite{Yuen1,Yuen2,EPRinSqueezing}. Conventionally, homodyne detection followed by an electrical intensity detector is used for measuring squeezing level as in Fig. \ref{concept}(a). In this scheme, correlated sidebands of the squeezed light at the frequency of $\Omega\pm\omega$ is down-converted into an electrical signal at $\omega$. Here, we consider generating a squeezed vacuum with an OPA and then measuring it with another OPA and an optical intensity detector, as in Fig. \ref{concept}(b). In this case, the optical intensity at $\Omega\pm\omega$ is directly measured without any down-conversion.

Let the gain of the second OPA be $G$. Photon number operator $\hat{n}$ of the light at the output of the second OPA at a frequency $\Omega + \omega$ is written as \cite{LiftingBandwidth}:
\begin{eqnarray}
    \hat{n}(\Omega+\omega) &=& \frac{G}{4}\hat{x}^{\dag}(\omega)\hat{x}(\omega) + \frac{1}{4G}\hat{p}^{\dag}(\omega)\hat{p}(\omega)-\frac{1}{2}, \\
  \hat{x}(\omega) & \equiv & \hat{a}(\Omega+\omega)+\hat{a}^{\dag}(\Omega-\omega), \\
  \hat{p}(\omega) & \equiv & i\hat{a}^{\dag}(\Omega+\omega)-i\hat{a}(\Omega-\omega),
\end{eqnarray}
where $\Omega$ is the center frequency of the OPAs, and annihilation operators $\hat{a}(\Omega\pm\omega)$ corresponds to the input for the second OPA, which is non-classically correlated sidebands of the squeezed vacuum generated in the first OPA. $\hat{x}^{\dag}(\omega)\hat{x}(\omega)$ and $\hat{p}^{\dag}(\omega)\hat{p}(\omega)$ represent the variances of the quadrature amplitudes of the squeezed vacuum in a rotating frame at the center frequency of $\Omega$. The variance varies between the squeezing level $R_{-}$ and the anti-squeezing level $R_{+}$ depending on the phase of the first OPA, $\theta$, as follows \cite{PhaseFluctuationEffect}:
\begin{eqnarray}
  \braket{\hat{x}^{\dag}(\omega)\hat{x}(\omega)} & = & R_{-}\cos^2\theta+R_{+}\sin^2\theta,\label{phasedepend1}\\
  \braket{\hat{p}^{\dag}(\omega)\hat{p}(\omega)} & = & R_{-}\sin^2\theta+R_{+}\cos^2\theta.\label{phasedepend2}
\end{eqnarray}
Here, we consider measuring the optical intensity at the frequency of $\Omega+\omega$ at the output of the cascaded OPAs by using an optical spectrum analyzer or a power meter with a color filter. The maximum and minimum values of the intensity $I_{max}$ and $I_{min}$ are obtained at $\theta=\pi/2,\pi$ as:
\begin{eqnarray}
  I_{max} & = & \frac{\hbar(\Omega+\omega)}{4}\left\{GR_{+}+\frac{R_{-}}{G}\right\}, \\
  I_{min} & = & \frac{\hbar(\Omega+\omega)}{4}\left\{GR_{-}+\frac{R_{+}}{G}\right\}.
\end{eqnarray}
Especially, when $R_{-}=R_{+}=1$, $I_{max}$ and $I_{min}$ are identical and the value corresponds to the intensity of the amplified vacuum
\begin{equation}
  I_{0} = \frac{\hbar(\Omega+\omega)}{4}\left(G+\frac{1}{G}\right).
\end{equation}
Normalized by the intensity of the amplified vacuum $I_0$, the measured squeezing level $R_{-}'$ is calculated as follows:
\begin{eqnarray}
  {R_{-}'} & \equiv & \frac{I_{min}}{I_{0}} =
  \frac{1}{1+G^2}R_{+}+\frac{G^2}{1+G^2}R_{-}.\label{measuredSQ}
\end{eqnarray}
Here, ${R_{-}'}$ decreases monotonically and approaches $R_{-}$ as $G$ increases. Compared to Eqs. \ref{phasedepend1} and \ref{phasedepend2}, the equation indicates that the effect of the finite gain is equivalent to phase deviation of
\begin{equation}
  \theta_{eff} = \arcsin \sqrt{\frac{1}{1+G^2}}. \label{effectivetheta}
\end{equation}
In conventional homodyne measurements, the typical value of the phase deviation is from $0.8^{\circ}$ \cite{OPO:Serikawa} to $4.3^{\circ}$ \cite{AokiPhase}. The larger anti-squeezing level is, the smaller acceptable phase deviation is. To know how large gain is needed for all-optical phase-sensitive detection, we consider two cases for example. The first case is -3.0 dB squeezing with 3.0 dB anti-squeezing, where 13 dB ($G=20$) of gain is enough to measure the squeezing level correctly in the significant digit. The second case is -3.0 dB squeezing with 15.0 dB anti-squeezing, where 19 dB ($G=80$) of gain is required.

In addition, when gain is finite, the actual squeezing level is obtained by following equations:
\begin{eqnarray}
  {R_{-}} & = & \frac{G^2}{G^2-1}{R_{-}'}-\frac{1}{G^2-1}{R_{+}'},\label{equation1} \\
  {R_{+}} & = & -\frac{1}{G^2-1}{R_{-}'}+\frac{G^2}{G^2-1}{R_{+}'},\label{equation2}
\end{eqnarray}
where ${R_{+}'}=I_{max}/I_0$. This correction formula might be useful in the measurement of squeezing level with a low-gain bulk-crystal OPA such as \cite{LossToleranceOfSU11}.
%%%%%%%%%%%%%%%%%%%%%%%%%%%%%%%%%%%%%%%%%%%%%%%%%%%%%%%%%%%%%%%%%%%%%%
\section{Experimental Setup}
Figure \ref{setup} shows a schematic of our experimental setup. A source of continuous-wave laser light at 1545.3 nm (194 THz) is a narrow-linewidth and low-noise seed laser (NKT Photonics, BASIK module). The output of the seed laser is split by 3-dB fiber coupler. One of the outputs of the coupler is amplified by a Erbium-doped fiber amplifier (Keopsys, CEFA-C-PB-HP), and the amplified beam passes through an optical band pass filter (Alnair Labs, TFF-15-1-PM-L-100-FS) to reduce noise from the fiber amplifier. Then, it pumps a frequency doubler (NTT Electronics, WH-0772-000-F-B-C). The frequency-doubled beam pumps a pigtaled PPLN waveguide OPA module OPA1, which is assembled similarly as in \cite{4dBFiberedOPA}, and generates a squeezed vacuum. The intensity of the frequency-doubled beam is monitored at the transmittance port of OPA1 by a photo detector PD1 (Newport, 818-SL). 
The other output of the 3-dB coupler is phase-controlled by a phase modulator (Covega, Mach-10) and then amplified by another Erbium-doped fiber amplifier (Keopsys, CEFA-C-PB-HP). The amplified beam passes through an optical band pass filter (Alnair Labs, TFF-15-1-PM-L-100-SS-SA), and then the amplified beam pumps another frequency doubler (NTT Electronics, WH-0772-000-F-B-C). The frequency-doubled beam pumps another pigtaled OPA module OPA2. The intensity of the frequency doubled beam is monitored at the transmittance port of OPA2 by a photo detector PD2 (Newport, 818-SL).

OPA1 and OPA2 are connected with their 1-m polarization-maintaining optical fiber pigtales (Fujikura, SM15-PS-U25D), and a quadrature amplitude of the squeezed vacuum from OPA1 is amplified in OPA2. We call the output of OPA2 at 1.5 $\mu$m "amplified squeezed vacuum." The output is split by a 10-dB coupler, and its main output is injected into an optical spectrum analyzer (Advantest, Q8384). The other output passes through a variable band-pass filter (santec, OTF-350) and its power and wavelength are measured by a photo detector PD3 (Thorlabs, PDA10CS2) and another optical spectrum analyzer (Anritsu, MS9710C) used as a wavelength monitor.

The signal from PD3 is processed by a field-programmable gate array (FPGA) board (Red Pitaya, STEMLAB 125-14). In the FPGA board, the signal is numerically subtracted from a target value and then time-integrated. The time-integrated signal from the FPGA board drives the phase modulator in the optical path. This feedback loop controls the value detected by PD3 to approach a target value. An integral control circuit locks signal not at peaks but on slope. However, since the phase differs depending on the wavelength due to chromatic dispersion, it is possible to lock the phase at any point by operating the variable optical bandpass filter. Phase control of parametric amplification of squeezed light would be not only indispensable for all-optical quantum computing, but also useful for quantum meteorology. For instance, the technique could be applied for a nonlinear interferometer \cite{SU(2)andSU(11),NonlinearInterferometers} whose phase is currently not locked but scanned as in \cite{Gaetano}.

\section{Result and Discussion}
\begin{figure}[!t]
\centering\includegraphics[width=8.6cm]{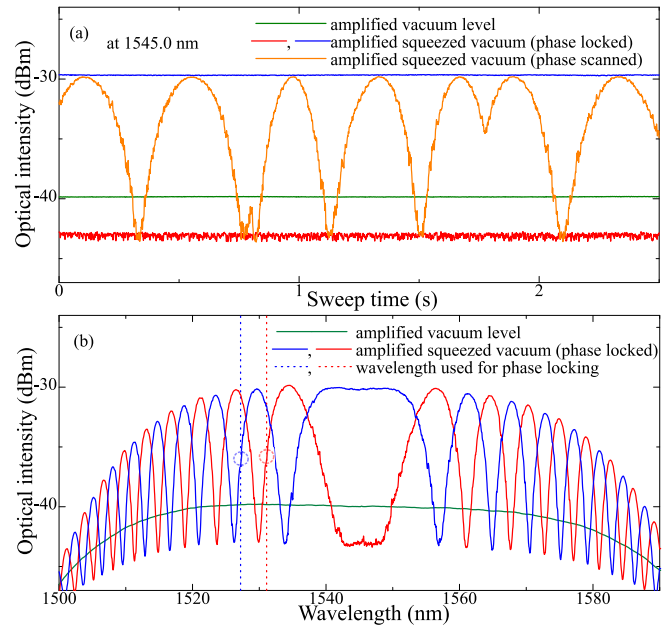}
\caption{The result of all-optical phase-sensitive detection obtained with an optical spectrum analyzer in (a) a zero-span mode at 1545.0 nm and (b) a range of wavelength from 1500 nm to 1590 nm. The resolution bandwidth is set to be 0.2 nm, and the smoothing-window width is (a) 2 ms, (b) 0.1 nm. The green curve is the optical spectrum of an amplified vacuum. Blue and red curves are those of a phase-locked amplified squeezed vacuum. Orange curve is phase-scanned signal of the amplified squeezed vacuum at 1545.0 nm.  The intensity of the pump beam for OPA1 measured at PD1 is 100 mW and the gain of second OPA is 23 dB.}
\label{blue}
\end{figure}
\begin{figure}[!t]
\centering\includegraphics[width=8cm]{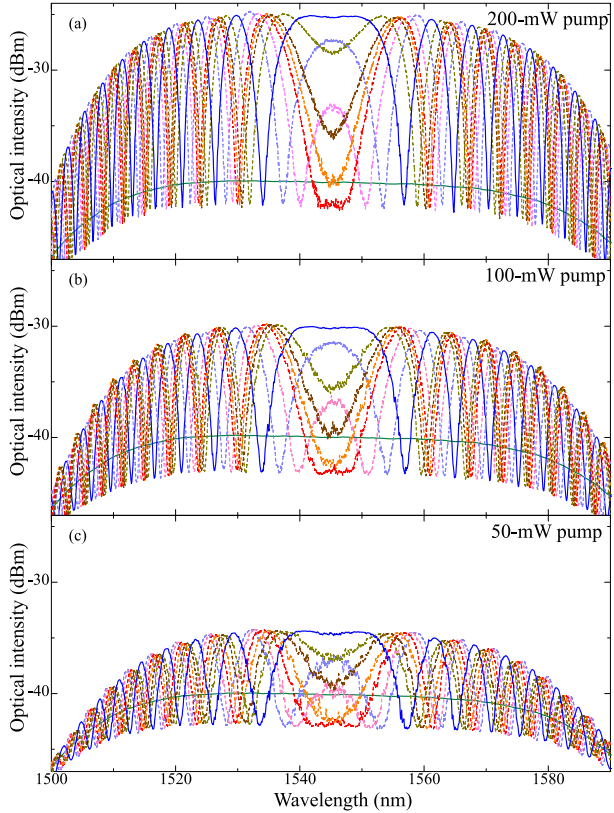}
\caption{Optical spectra of phase-locked amplified squeezed vacua with various locking phases and pump powers. That of an amplified vacuum is indicated by green curve. The resolution bandwidth is set to be 0.2 nm, and the smoothing-window width is 0.1 nm. Green curve is the optical spectrum of an amplified vacuum. Other colored curves are those of an amplified squeezed vacuum at different phase-locking points. The intensity of the pump beam for OPA1 measured at PD1 is (a) 200 mW, (b) 100 mW, and (c) 50 mW.}
\label{param}
\end{figure}
\subsection{Phase-locked all-optical phase-sensitive detection}
Figure \ref{blue} (a) shows the optical signal of all-optical phase-sensitive detection at 1545.0 nm. The pump power for OPA1 measured at PD1 is 100 mW, and that for OPA2 measured at PD2 is 300 mW. The green and orange curves are the intensity of amplified vacuum and squeezed vacuum with scanned phase, respectively. The blue and red curves are those with locked phases. Figure \ref{blue} (b) shows the spectra of the phase-locked amplified squeezed vacuum with the spectrum of the amplified vacuum. Although the phase is locked for each wavelength, the phase of the amplified squeezed vacuum depends on the wavelength due to the chromatic dispersion of the 2-m PM fiber between the OPAs. Here, wavelengths indicated by dashed lines are selected by the variable bandpass filter and used as an error signal for phase locking. The intersection of the solid curves and dashed lines indicated by a dashed circle is placed on the middle of the ripple, which corresponds to the target value set in the FPGA board.

The parametric gain of OPA2 with 300-mW pump is measured to be 23 dB, namely 200, in advance. Assigning the gain into Eq. \ref{effectivetheta}, the effective phase deviation for the gain is $0.3^{\circ}$ and is negligible within the significant digits of squeezing and anti-squeezing levels, which means the gain of 23 dB is enough for all-optical phase-sensitive detection of the squeezed light.

Figure \ref{param} shows the spectrum for various pump powers and phase-locking points. The phase-locking point is changed by manually rotating a micrometer of the variable bandpass filter. The squeezing level and anti-squeezing level depend on the pump power for OPA1, and those around the center frequency are measured to be -2.2 dB and 14.9 dB at 200-mW pump, -3.2 dB and 9.9 dB at 100-mW pump, and -2.7 dB and 5.6 dB at 50-mW pump, respectively. Thanks to the flat amplification characteristic of the OPA from 1520 nm (197 THz) to 1570 nm (191 THz), squeezing at sideband frequencies up to 3 THz is well observed at 100-mW pump. The slight decrease in the squeezing level at 200-mW pump is considered due to contamination with the large anti-squeezing components caused by residual phase noise from the light source and imperfections of wavelength filtering in the optical spectrum analyzer.
\begin{figure}[!t]
\centering\includegraphics[width=7cm]{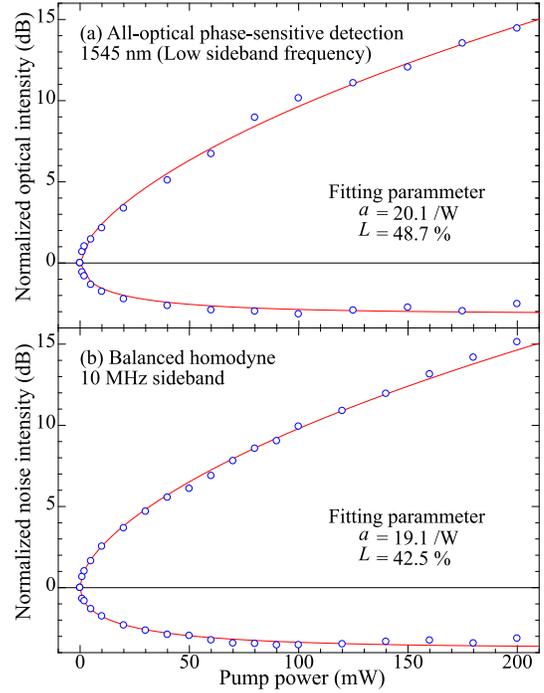}
\caption{Pump power dependency of squeezing and anti-squeezing level obtained by (a) the all-optical phase-sensitive detection and (b) a conventional balanced homodyne measurement. The pump power is measured at PD1. The results of the two measurement methods are in good agreement.}
\label{fitting}
\end{figure}

\subsection{Comparison with balanced homodyne measurement}
To compare with the all-optical phase-sensitive detection, we also perform measurement of a squeezed vacuum with a homemade balanced homodyme detector with InGaAs photodiodes (Laser Components, IGHQEX0100-1550-10-1.0-SPAR-TH-40), which was also used in \cite{4dBFiberedOPA}. To match the conditions, the phase is not locked but scanned as in Fig. \ref{blue}(a). In the balanced homodyne measurement, the squeezed vacuum is interfered with a 2.5-mW local oscillator beam in a fiber beamsplitter (Thorlabs, PN1550R5F2) spliced with AR-coated fiber (P1-1550PMAR-2). The electrical signal from the detector is measured with an electrical spectrum analyzer (Keysight, N9010B). The resolution bandwidth, video bandwidth, and analysis frequency are set to be 3 MHz, 1 kHz, and 10 MHz, respectively.

Figure \ref{fitting} shows the pump power dependency of squeezing and anti-squeezing level obtained by two measurement methods. The squeezing and anti-squeezing level are described as \cite{SqPhaseError}:
\begin{eqnarray}
  R_{\pm} &=& L + (1-L) e^{\pm 2\sqrt{ap}}
\end{eqnarray}
Here, $a$ is nonlinear efficiency of an OPA; $p$ is the intensity of the pump beam for the OPA; $L$ is total optical loss. The efficiency $a$ and loss $L$ are fitted to be 19.1 /W and 42.5\% for the balanced homodyne measurement. Considering the excess loss of 14\% in the fiber beamsplitter, the effective loss of 2\% due to circuit noise, and the quantum efficiency of 93\% in the detector including collimating and focusing lenses, the total detection efficiency in the balanced homodyne detection setup is calculated to be 78\%, and the optical loss of a squeezed vacuum in OPA1 is estimated to be 27\%.

The efficiency $a$ and loss $L$ are fitted to be 20.1 /W and 48.7\% for the all-optical phase-sensitive detection. The "quantum efficiency" of the OPA2 as a detector is calculated to be $(1-0.487)/(1-0.27)\approx0.70$, which is considered to be the loss in OPA2 and the fiber joint. 
The loss could be reduced by improving the coupling efficiency between the waveguide and the fiber in the OPA module and also by reducing the propagation loss in the waveguide due to the surface roughness \cite{4dBFiberedOPA}.
\subsection{Dispersion compensation}
\begin{figure}[!t]
\centering\includegraphics[width=8.5cm]{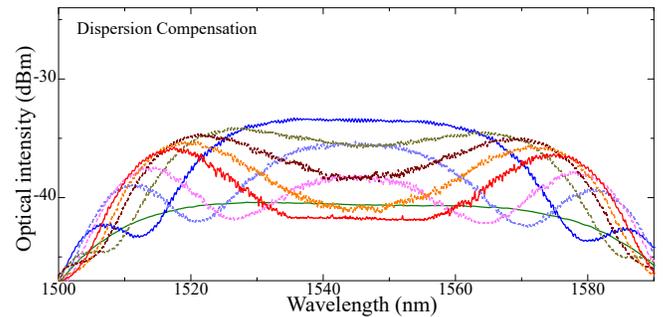}
\caption{The result of dispersion compensation of a phase-locked squeezed vacuum measured by the optical spectrum analyzer. The setup and settings are same as Figs. \ref{setup} and \ref{param} except the insertion of a dispersion-compensating fiber. The intensity of the pump beam for OPA1 measured at PD1 is 100 mW.}
\label{DCF}
\end{figure}
In the spectrum shown in Fig. \ref{param}, there are ripples due to the chromatic dispersion of the fibers between the OPAs. 
We also demonstrate the compensation of the dispersion of the squeezed vacuum. In the researches of telecommunication, it is known that the 2nd-order chromatic dispersion $D$ relates to the gain spectrum $g(f)$ of cascaded OPAs as \cite{DispersionAsobe,ShimizuDispersion}:
\begin{eqnarray}
  g(f) &=& g_0 \cos^2 \left(\phi(f)\right) + g_0^{-1} \sin^2 \left(\phi(f)\right), \label{gainspectrum}\\
  \phi (f) &=& \pi D c \left(\frac{f_0 - f}{f_0}\right)^2 +\phi_0,
\end{eqnarray}
where $f$ is frequency of light; $f_0$ is the center frequency, namely 194 THz;
$c$ is the speed of light; $g_0$ is the gain of second OPA; $\phi$ is the phase corresponding to the locking point. Equation \ref{gainspectrum} is modified for the measurement of squeezed light as follows:
\begin{eqnarray}
  R(f) &=& R_{+} \cos^2 \left(\phi(f)\right) + R_{-} \sin^2 \left(\phi(f)\right),
\end{eqnarray}
where $R(f)$ is the measured spectrum normalized by the amplified vacuum level. By using the equation, we estimate the dispersion $D$ for the spectra in Fig. \ref{param} to be 0.033 ps/nm, which is reasonable value as the dispersion of the 2-m single-mode PM optical fiber \cite{DispersionValue}. 

To eliminate the dispersion, a dispersion compensating fiber (DCF) patch cable (assembled by Optronscience inc.) is inserted between the optical fiber pigtales of the two OPAs. The DCF patch cable consists of 70-cm DCF (Thorlabs, PMDCF) spliced to 15-cm PM optical fibers (Thorlabs, PM1550-XP) at both ends and SC/PC connectors, and the length of the fibers is designed to counteract the dispersion of the 2-m PM fiber. Figure \ref{DCF} shows the result of all-optical phase-sensitive detection with the DCF. The interval of the ripple is lengthened, and it corresponds to the residual dispersion of 0.0045 ps/nm. The red curve in the Fig. \ref{DCF} shows that the phase of the squeezed light is maintained over 1 THz (from 1537 nm to 1545 nm), which is about twice as much as that without the DCF as shown in Fig. \ref{param} (from around 1541 nm to 1545 nm). Further reduction of chromatic dispersion could be achieved with a spatial light modulator as in \cite{ShimizuDispersion} or by integrating two OPAs into one LiNbO${}_{3}$ integrated optical circuit such as \cite{fiberOPA20dB0,integratedLN1,integratedLN2}. Additionally, the squeezing and anti-squeezing level, -1.2 dB and 7.1 dB around the center wavelength, are consistent with the insertion loss of the DCF, 2.9 dB. Squeezed light with locked phase maintained over 1 THz would allow a ``qumode \cite{QumodeVanLoock}'' to be defined in a micron-order wave packet in time-domain multiplexed quantum computation \cite{PPLNOPA6dB}.
\section{Conclusion}
We demonstrated all-optical phase-sensitive detection of a quadrature amplitude of squeezed light using a fiber-coupled PPLN waveguide OPA. Squeezing level of 3 dB is observed over 3 THz of sideband frequency. The measured squeezing level is consistent with that of conventional phase-sensitive detection, homodyne measurement. The phase of broadband squeezed light is locked with a variable optical bandpass filter, which enables to lock the phase to an arbitrary point. Furthermore, we performed the dispersion compensation of the broadband squeezed light, so that the phase of the squeezed light is maintained over 1 THz. Our work would help to realize all-optical quantum computation with over-THz clock frequency.

\section*{Funding Information}
This work is funded by Core Research for Evolutional Science and Technology (CREST) (JPMJCR15N5) of Japan Science and Technology Agency (JST), KAKENHI (18H05207) of Japan Society for the Promotion of Science (JSPS), APLS of Ministry of Education, Culture, Sports, Science and Technology (MEXT), and The University of Tokyo Foundation.

\section*{Disclosures}
The authors declare no conflicts of interest.
%\section*{References}
%For references, you may add citations manually or use BibTeX. E.g. \cite{Zhang:14}.
%Letter submissions to \emph{Optica} require two sets of references: an abbreviated reference style for publication and a full reference list to aid the editor and reviewers. Citations to journal articles in the abbreviated list should omit the article title and final page number; this abbreviated reference style is produced automatically when the \texttt{$\setminus$setboolean\{shortarticle\}\{true\}} option is selected in the template, if you are using a .bib file for your references.
 %The full reference list meant to aid the editor and reviewers must be included as well on an informational page that will not count against page length; again this will be produced automatically if you are using a .bib file and have the \texttt{$\setminus$setboolean\{shortarticle\}\{true\}} option selected.
% Bibliography
\bibliography{sample}

\begin{thebibliography}{10}
\newcommand{\enquote}[1]{``#1''}

\bibitem{twoDcluster2}
M.~V. Larsen, X.~Guo, C.~R. Breum, J.~S. Neergaard-Nielsen, and U.~L. Andersen,
  \enquote{Deterministic generation of a two-dimensional cluster state,}
  {\protect\JournalTitle{Science}} \textbf{366}, 369--372 (2019).

\bibitem{twoDcluster}
W.~Asavanant, Y.~Shiozawa, S.~Yokoyama, B.~Charoensombutamon, H.~Emura, R.~N.
  Alexander, S.~Takeda, J.-i. Yoshikawa, N.~C. Menicucci, H.~Yonezawa, and
  A.~Furusawa, \enquote{Generation of time-domain-multiplexed two-dimensional
  cluster state,} {\protect\JournalTitle{Science}} \textbf{366}, 373--376
  (2019).

\bibitem{OneWay}
R.~Raussendorf and H.~J. Briegel, \enquote{A one-way quantum computer,}
  {\protect\JournalTitle{Phys. Rev. Lett.}} \textbf{86}, 5188--5191 (2001).

\bibitem{MillionModes:Yoshikawa}
J.~Yoshikawa, S.~Yokoyama, T.~Kaji, C.~Sornphiphatphong, Y.~Shiozawa,
  K.~Makino, and A.~Furusawa, \enquote{Invited article: Generation of
  one-million-mode continuous-variable cluster state by unlimited time-domain
  multiplexing,} {\protect\JournalTitle{APL Photonics}} \textbf{1}, 060801
  (2016).

\bibitem{PPLNOPA6dB}
T.~Kashiwazaki, N.~Takanashi, T.~Yamashima, T.~Kazama, K.~Enbutsu, R.~Kasahara,
  T.~Umeki, and A.~Furusawa, \enquote{Continuous-wave 6-{dB}-squeezed light
  with 2.5-{THz}-bandwidth from single-mode {PPLN} waveguide,}
  {\protect\JournalTitle{submitted to APL Photonics}}  (2019).

\bibitem{alloptical}
T.~C. Ralph, \enquote{All-optical quantum teleportation,}
  {\protect\JournalTitle{Optics letters}} \textbf{24}, 348--350 (1999).

\bibitem{FurusawaTeleportation}
A.~Furusawa, J.~L. S{\o}rensen, S.~L. Braunstein, C.~A. Fuchs, H.~J. Kimble,
  and E.~S. Polzik, \enquote{Unconditional quantum teleportation,}
  {\protect\JournalTitle{Science}} \textbf{282}, 706--709 (1998).

\bibitem{Yuen1}
H.~P. Yuen and J.~H. Shapiro, \enquote{Generation and detection of two-photon
  coherent states in degenerate four-wave mixing,} {\protect\JournalTitle{Opt.
  Lett.}} \textbf{4}, 334--336 (1979).

\bibitem{Yuen2}
H.~P. Yuen and V.~W.~S. Chan, \enquote{Noise in homodyne and heterodyne
  detection,} {\protect\JournalTitle{Opt. Lett.}} \textbf{8}, 177--179 (1983).

\bibitem{EPRinSqueezing}
J.~Zhang, \enquote{Einstein-podolsky-rosen sideband entanglement in broadband
  squeezed light,} {\protect\JournalTitle{Phys. Rev. A}} \textbf{67}, 054302
  (2003).

\bibitem{GHzSq:Schnabel}
S.~Ast, M.~Mehmet, and R.~Schnabel, \enquote{High-bandwidth squeezed light at
  1550 nm from a compact monolithic {PPKTP} cavity,}
  {\protect\JournalTitle{Opt. Express}} \textbf{21}, 13572--13579 (2013).

\bibitem{FastHomodyne1}
F.~Raffaelli, G.~Ferranti, D.~H. Mahler, P.~Sibson, J.~E. Kennard,
  A.~Santamato, G.~Sinclair, D.~Bonneau, M.~G. Thompson, and J.~C.~F. Matthews,
  \enquote{A homodyne detector integrated onto a photonic chip for measuring
  quantum states and generating random numbers,} {\protect\JournalTitle{Quantum
  Science and Technology}} \textbf{3}, 025003 (2018).

\bibitem{FastHomodyne2}
X.~{Zhang}, Y.~{Zhang}, Z.~{Li}, S.~{Yu}, and H.~{Guo}, \enquote{1.2-{GHz}
  balanced homodyne detector for continuous-variable quantum information
  technology,} {\protect\JournalTitle{IEEE Photonics Journal}} \textbf{10},
  1--10 (2018).

\bibitem{serikawatheory}
T.~Serikawa and A.~Furusawa, \enquote{Excess loss in homodyne detection
  originating from distributed photocarrier generation in photodiodes,}
  {\protect\JournalTitle{Phys. Rev. Applied}} \textbf{10}, 064016 (2018).

\bibitem{serikawadetector}
T.~Serikawa and A.~Furusawa, \enquote{500 {MHz} resonant photodetector for
  high-quantum-efficiency, low-noise homodyne measurement,}
  {\protect\JournalTitle{Review of Scientific Instruments}} \textbf{89}, 063120
  (2018).

\bibitem{BroadbandOPA}
Z.~{Tong}, C.~{Lundstr\"{o}m}, P.~A. {Andrekson}, M.~{Karlsson}, and
  A.~{Bogris}, \enquote{Ultralow noise, broadband phase-sensitive optical
  amplifiers, and their applications,} {\protect\JournalTitle{IEEE Journal of
  Selected Topics in Quantum Electronics}} \textbf{18}, 1016--1032 (2012).

\bibitem{FiberBroadbandOPA}
M.~E. Marhic, N.~Kagi, T.-K. Chiang, and L.~G. Kazovsky, \enquote{Broadband
  fiber optical parametric amplifiers,} {\protect\JournalTitle{Opt. Lett.}}
  \textbf{21}, 573--575 (1996).

\bibitem{WaveguideBroadbandOPA}
T.~Umeki, M.~Asobe, T.~Yanagawa, O.~Tadanaga, Y.~Nishida, K.~Magari, and
  H.~Suzuki, \enquote{Broadband wavelength conversion based on apodized
  $\chi^{(2)}$ grating,} {\protect\JournalTitle{J. Opt. Soc. Am. B}}
  \textbf{26}, 2315--2322 (2009).

\bibitem{OPAFurusawa}
K.~Yoshino, T.~Aoki, and A.~Furusawa, \enquote{Generation of continuous-wave
  broadband entangled beams using periodically poled lithium niobate
  waveguides,} {\protect\JournalTitle{Applied Physics Letters}} \textbf{90},
  041111 (2007).

\bibitem{YamamotoQND}
Y.~Yamamoto and H.~A. Haus, \enquote{Preparation, measurement and information
  capacity of optical quantum states,} {\protect\JournalTitle{Rev. Mod. Phys.}}
  \textbf{58}, 1001--1020 (1986).

\bibitem{NoiseFactorOfAmplifiers}
C.~M. Caves, \enquote{Quantum limits on noise in linear amplifiers,}
  {\protect\JournalTitle{Phys. Rev. D}} \textbf{26}, 1817--1839 (1982).

\bibitem{NoiseFactorOfAmplifiers2}
H.~A. Haus and J.~A. Mullen, \enquote{Quantum noise in linear amplifiers,}
  {\protect\JournalTitle{Phys. Rev.}} \textbf{128}, 2407--2413 (1962).

\bibitem{SU(2)andSU(11)}
B.~Yurke, S.~L. McCall, and J.~R. Klauder, \enquote{Su(2) and {SU}(1,1)
  interferometers,} {\protect\JournalTitle{Phys. Rev. A}} \textbf{33},
  4033--4054 (1986).

\bibitem{EntangledetectionbyOPA}
J.~Li, Y.~Liu, N.~Huo, L.~Cui, S.~Feng, X.~Li, and Z.~Y. Ou, \enquote{Measuring
  continuous-variable quantum entanglement with parametric-amplifier-assisted
  homodyne detection,} {\protect\JournalTitle{Phys. Rev. A}} \textbf{101},
  053801 (2020).

\bibitem{LossToleranceOfSU11}
M.~Manceau, G.~Leuchs, F.~Khalili, and M.~Chekhova, \enquote{Detection loss
  tolerant supersensitive phase measurement with an {SU}(1,1) interferometer,}
  {\protect\JournalTitle{Phys. Rev. Lett.}} \textbf{119}, 223604 (2017).

\bibitem{allopticalexperiment}
S.~Liu, Y.~Lou, and J.~Jing, \enquote{Orbital angular momentum multiplexed
  deterministic all-optical quantum teleportation,}
  {\protect\JournalTitle{Nature Communications}} \textbf{11}, 1--8 (2020).

\bibitem{Kashiwazaki}
T.~Kashiwazaki, K.~Enbutsu, T.~Kazama, O.~Tadanaga, T.~Umeki, and R.~Kasahara,
  \enquote{Over-30-d{B} phase-sensitive amplification using a fiber-pigtailed
  {PPLN} waveguide module,} in \emph{Nonlinear Optics (NLO),}  (Optical Society
  of America, 2019), p. NW3A.2.

\bibitem{LiftingBandwidth}
Y.~Shaked, Y.~Michael, R.~Z. Vered, L.~Bello, M.~Rosenbluh, and A.~Pe’er,
  \enquote{Lifting the bandwidth limit of optical homodyne measurement with
  broadband parametric amplification,} {\protect\JournalTitle{Nature
  communications}} \textbf{9}, 1--12 (2018).

\bibitem{4dBFiberedOPA}
N.~{Takanashi}, T.~{Kashiwazaki}, T.~{Kazama}, K.~{Enbutsu}, R.~{Kasahara},
  T.~{Umeki}, and A.~{Furusawa}, \enquote{4-{dB} quadrature squeezing with
  fiber-coupled ppln ridge waveguide module,} {\protect\JournalTitle{IEEE
  Journal of Quantum Electronics}} \textbf{56}, 1--5 (2020).

\bibitem{PhaseFluctuationEffect}
T.~C. Zhang, K.~W. Goh, C.~W. Chou, P.~Lodahl, and H.~J. Kimble,
  \enquote{Quantum teleportation of light beams,} {\protect\JournalTitle{Phys.
  Rev. A}} \textbf{67}, 033802 (2003).

\bibitem{OPO:Serikawa}
T.~Serikawa, J.~Yoshikawa, K.~Makino, and A.~Frusawa, \enquote{Creation and
  measurement of broadband squeezed vacuum from a ring optical parametric
  oscillator,} {\protect\JournalTitle{Opt. Express}} \textbf{24}, 28383--28391
  (2016).

\bibitem{AokiPhase}
T.~Aoki, G.~Takahashi, and A.~Furusawa, \enquote{Squeezing at 946 nm with
  periodically poled {KTiOPO}${}_4$,} {\protect\JournalTitle{Opt. Express}}
  \textbf{14}, 6930--6935 (2006).

\bibitem{NonlinearInterferometers}
M.~V. Chekhova and Z.~Y. Ou, \enquote{Nonlinear interferometers in quantum
  optics,} {\protect\JournalTitle{Adv. Opt. Photon.}} \textbf{8}, 104--155
  (2016).

\bibitem{Gaetano}
G.~Frascella, E.~E. Mikhailov, N.~Takanashi, R.~V. Zakharov, O.~V. Tikhonova,
  and M.~V. Chekhova, \enquote{Wide-field {SU}(1,1) interferometer,}
  {\protect\JournalTitle{Optica}} \textbf{6}, 1233--1236 (2019).

\bibitem{SqPhaseError}
T.~Hirano, K.~Kotani, T.~Ishibashi, S.~Okude, and T.~Kuwamoto, \enquote{3 {dB}
  squeezing by single-pass parametric amplification in a periodically poled
  {KTiOPO}${}_{4}$ crystal,} {\protect\JournalTitle{Opt. Lett.}} \textbf{30},
  1722--1724 (2005).

\bibitem{DispersionAsobe}
M.~Asobe, T.~Umeki, K.~Enbutsu, O.~Tadanaga, and H.~Takenouchi, \enquote{Phase
  squeezing and dispersion tolerance of phase sensitive amplifier using
  periodically poled {LiNbO}${}_{3}$ waveguide,} {\protect\JournalTitle{J. Opt.
  Soc. Am. B}} \textbf{31}, 3164--3169 (2014).

\bibitem{ShimizuDispersion}
S.~Shimizu, T.~Kazama, T.~Kobayashi, T.~Umeki, K.~Enbutsu, R.~Kasahara, and
  Y.~Miyamoto, \enquote{Gain ripple and passband narrowing due to residual
  chromatic dispersion in non-degenerate phase-sensitive amplifiers,} in
  \emph{Optical Fiber Communication Conference (OFC) 2020,}  (Optical Society
  of America, 2020), p. M1I.3.

\bibitem{DispersionValue}
K.~Okamoto and T.~Hosaka, \enquote{Polarization-dependent chromatic dispersion
  in birefringent optical fibers,} {\protect\JournalTitle{Opt. Lett.}}
  \textbf{12}, 290--292 (1987).

\bibitem{fiberOPA20dB0}
F.~Mondain, T.~Lunghi, A.~Zavatta, E.~Gouzien, F.~Doutre, M.~D. Micheli,
  S.~Tanzilli, and V.~D'Auria, \enquote{Chip-based squeezing at a telecom
  wavelength,} {\protect\JournalTitle{Photon. Res.}} \textbf{7}, A36--A39
  (2019).

\bibitem{integratedLN1}
G.~S. Kanter, P.~Kumar, R.~V. Roussev, J.~Kurz, K.~R. Parameswaran, and M.~M.
  Fejer, \enquote{Squeezing in a {LiNbO}${}_{3}$ integrated optical waveguide
  circuit,} {\protect\JournalTitle{Opt. Express}} \textbf{10}, 177--182 (2002).

\bibitem{integratedLN2}
C.~Wang, M.~Zhang, X.~Chen, M.~Bertrand, A.~Shams-Ansari, S.~Chandrasekhar,
  P.~Winzer, and M.~Lon{\v{c}}ar, \enquote{Integrated lithium niobate
  electro-optic modulators operating at cmos-compatible voltages,}
  {\protect\JournalTitle{Nature}} \textbf{562}, 101--104 (2018).

\bibitem{QumodeVanLoock}
P.~van Loock, \enquote{Optical hybrid approaches to quantum information,}
  {\protect\JournalTitle{Laser \& Photonics Reviews}} \textbf{5}, 167--200
  (2011).

\end{thebibliography}
% Full bibliography will be added automatically on a new page for Optics Letters submissions. This command is ignored for journal article submissions.
% Note that this extra page will not count against page length.
\bibliographyfullrefs{sample}

%Manual citation list
%\begin{thebibliography}{1}
%\bibitem{Zhang:14}
%Y.~Zhang, S.~Qiao, L.~Sun, Q.~W. Shi, W.~Huang, %L.~Li, and Z.~Yang,
 % \enquote{Photoinduced active terahertz metamaterials with nanostructured
  %vanadium dioxide film deposited by sol-gel method,} Opt. Express \textbf{22},
  %11070--11078 (2014).
%\end{thebibliography}

\end{document}